\documentclass[%
 reprint,
 superscriptaddress,
 amsmath,amssymb,
 aps,
]{revtex4-1}

\usepackage{graphicx}
 \usepackage{dcolumn}
\usepackage{bm}
\usepackage{color}
\usepackage[normalem]{ulem}

\usepackage{hyperref}

\begin{document}

\preprint{ZU-TH-30/15}

\title{Global constraints on anomalous triple gauge couplings \\ 
in effective field theory approach
}

\author{Adam Falkowski}
\affiliation{%
Laboratoire de Physique Th\'{e}orique, Bat.~210, Universit\'{e} Paris-Sud, 91405 Orsay, France
}
\author{Mart\'{i}n Gonz\'{a}lez-Alonso}%
\affiliation{%
IPN de Lyon/CNRS, Universite Lyon 1, Villeurbanne, France
}%
\author{Admir Greljo}
\affiliation{Physik-Institut, Universitat Z\"{u}rich, CH-8057 Z\"{u}rich, Switzerland}
\affiliation{%
Faculty of Science, University of Sarajevo, Zmaja od Bosne 33-35, 71000 Sarajevo, Bosnia and Herzegovina
}%
\author{David Marzocca}
\affiliation{Physik-Institut, Universitat Z\"{u}rich, CH-8057 Z\"{u}rich, Switzerland}



\begin{abstract}

We present a combined analysis of 
LHC Higgs data (signal strengths)
together with LEP-2 WW production measurements. 
To characterize possible deviations from the Standard Model (SM) predictions,  we employ the framework of an Effective Field Theory (EFT) where the SM is extended by higher-dimensional operators suppressed by the mass scale of new physics $\Lambda$. 
The analysis is performed consistently at the order $\Lambda^{-2}$ in the EFT expansion keeping all the relevant operators. 
While the two data sets suffer from flat directions, together they impose stringent model-independent constraints on the anomalous triple gauge couplings.
As a side product, we provide the results of the combined fit in different EFT bases.

\end{abstract}

\pacs{Valid PACS appear here}

\maketitle

\newcommand{\fref}[1]{Fig.~\ref{fig:#1}} 
\newcommand{\eref}[1]{Eq.~\eqref{eq:#1}} 
\newcommand{\erefn}[1]{ (\ref{eq:#1})}
\newcommand{\erefs}[2]{Eqs.~(\ref{eq:#1}) - (\ref{eq:#2}) } 
\newcommand{\aref}[1]{Appendix~\ref{app:#1}}
\newcommand{\sref}[1]{Section~\ref{sec:#1}}
\newcommand{\cref}[1]{Chapter~\ref{ch:.#1}}
\newcommand{\tref}[1]{Table~\ref{tab:#1}}

\newcommand{\nn}{\nonumber \\}  
\newcommand{\nnl}{\nonumber \\}  
\newcommand{\nl}{& \nonumber \\ &}
\newcommand{\bnl}{\right .  \nonumber \\  \left .}
\newcommand{\dbnl}{\right .\right . & \nonumber \\ & \left .\left .}

\newcommand{\beq}{\begin{equation}} 
\newcommand{\eeq}{\end{equation}} 
\newcommand{\ba}{\begin{array}}  
\newcommand{\ea}{\end{array}} 
\newcommand{\bea}{\begin{eqnarray}}  
\newcommand{\eea}{\end{eqnarray} }  
\newcommand{\be}{\begin{eqnarray}}  
\newcommand{\ee}{\end{eqnarray} }  
\newcommand{\bal}{\begin{align}}
\newcommand{\eal}{\end{align}}   
\newcommand{\bi}{\begin{itemize}}  
\newcommand{\ei}{\end{itemize}}  
\newcommand{\ben}{\begin{enumerate}}  
\newcommand{\een}{\end{enumerate}}  
\newcommand{\bc}{\begin{center}}
\newcommand{\ec}{\end{center}} 
\newcommand{\bt}{\begin{table}}
\newcommand{\et}{\end{table}}  
\newcommand{\btb}{\begin{tabular}}
\newcommand{\etb}{\end{tabular}}  
\newcommand{\bvec}{\left ( \ba{c}}
\newcommand{\evec}{\ea \right )}

\newcommand{\cO}{{\mathcal O}} 
\newcommand{\co}{{\mathcal O}} 
\newcommand{\cL}{{\mathcal L}} 
\newcommand{\cl}{{\mathcal L}} 
\newcommand{\cM}{{\mathcal M}}

\newcommand{\const}{\mathrm{const}}

\newcommand{\ev}{ \mathrm{eV}}
\newcommand{\kev}{\mathrm{keV}}
\newcommand{\mev}{\mathrm{MeV}}
\newcommand{\gev}{\mathrm{GeV}}
\newcommand{\tev}{\mathrm{TeV}}

\newcommand{\mpl}{M_{\mathrm Pl}}

\def\mgut{\, M_{\rm GUT}}
\def\tgut{\, t_{\rm GUT}}
\def\mpl{\, M_{\rm Pl}}
\def\mkk{\, M_{\rm KK}}
\newcommand{\msusy}{M_{\rm soft}}

\newcommand{\dslash}[1]{#1 \! \! \! {\bf /}}
\newcommand{\ddslash}[1]{#1 \! \! \! \!  {\bf /}}

\def\ads{AdS$_5$\,}
\def\adse{AdS$_5$}
\def\intdk{\int {d^4 k \over (2 \pi)^4}} 

\def\ra{\rangle}
\def\la{\langle}  

\def\sgn{{\rm sgn}}
\def\pa{\partial}  
\newcommand{\dlr}{\overleftrightarrow{\partial}}
\newcommand{\Dlr}{\overleftrightarrow{D}}
\newcommand{\re}{{\mathrm{Re}} \,}
\newcommand{\im}{{\mathrm{Im}} \,}
\newcommand{\tr}{\mathrm T \mathrm r}  

\newcommand{\Ra}{\Rightarrow}
\newcommand{\lra}{\leftrightarrow}
\newcommand{\llra}{\longleftrightarrow}

\newcommand\simlt{\stackrel{<}{{}_\sim}}
\newcommand\simgt{\stackrel{>}{{}_\sim}}   
\newcommand{\zt}{$\mathbb Z_2$ }

\newcommand{\ha}{{\hat a}}
\newcommand{\hab}{{\hat b}}
\newcommand{\hac}{{\hat c}} 

\newcommand{\ti}{\tilde}  
\def\hc{{\rm h.c.}} 
\def\ov{\overline}  
  
\newcommand{\eps}{\epsilon}

\def\cog{\color{OliveGreen}}
\def\cor{\color{Red}}
\def\copu{\color{purple}}
\def\coro{\color{RedOrange}}
\def\coma{\color{Maroon}}
\def\cob{\color{Blue}}
\def\cobr{\color{Brown}}
\def\cobl{\color{Black}}
\def\cost{\color{WildStrawberry}}

The non-abelian local symmetry of the Standard Model (SM) implies that  cubic and quartic self-interactions of the gauge bosons must  be present in the Lagrangian.
An especially interesting example is the cubic interactions of W bosons with a photon or a Z boson because they can be directly probed in high-energy colliders such as LEP-2, Tevatron and the LHC.   
The SM uniquely predicts  the tensor structure of these interactions and fixes their  strength in terms the electromagnetic coupling $e$ and the weak mixing angle $\sin \theta_W \equiv s_\theta$.  

It has been recognized long ago that these predictions can be affected by new physics beyond the SM.  
This question can be addressed in a model-independent way in the linear EFT framework, i.e. with the Higgs field embedded in an $SU(2)$ doublet. 
In this approach, the SM is extended  by non-renormalizable gauge-invariant operators with  mass dimensions $D > 4$, 
which  encode the effects of new particles with  the mass scale $\Lambda$ much larger than the W boson mass $m_W$. 
The EFT approach organizes the new physics effects as an expansion in $1/\Lambda$, and the leading lepton-number-conserving corrections are $\cO(\Lambda^{-2})$ originating from $D$=6 operators. 
In the presence of {$D$=6 CP-conserving} operators,  the cubic couplings of electroweak gauge bosons take the form \cite{DeRujula:1991se,Hagiwara:1993ck}:
\bea  & 
\label{eq:atgc}
 \cL_{\rm tgc}  = 
i  e    \left ( W_{\mu \nu}^+ W_\mu^-  -  W_{\mu \nu}^- W_\mu^+ \right ) A_\nu  
\nnl &   
  +  i  e {c_\theta \over s_\theta} 
 \left (1 + \delta g_{1,z} \right )   \left ( W_{\mu \nu}^+ W_\mu^-  -  W_{\mu \nu}^- W_\mu^+ \right ) Z_\nu 
\nnl & 
+ i e (1 + \delta \kappa_\gamma)  A_{\mu\nu}\,W_\mu^+W_\nu^-   +   i  e {c_\theta \over s_\theta}  \left (1 +  \delta \kappa_z \right ) Z_{\mu\nu}\,W_\mu^+W_\nu^- 
  \nnl   & +     
 i   { \lambda_z  e  \over m_W^2 } \left [   W_{\mu \nu}^+W_{\nu \rho}^- A_{\rho \mu}  +  {c_\theta \over s_\theta} W_{\mu \nu}^+W_{\nu \rho}^- Z_{\rho \mu}   \right], 
\eea 
where $\delta \kappa_z  = \delta g_{1,z} -{s_\theta^2 \over c_\theta^2} \delta \kappa_\gamma$, and $c_\theta = \sqrt{1 - s_\theta^2}$. 
Therefore, as long as operators with $D >6$ are negligible, deformations of the  cubic gauge interactions due to new physics can be  parametrized by 3 anomalous triple gauge couplings (aTGCs): $\delta g_{1,z}$, $\delta \kappa_\gamma$, and $\lambda_z$.  
In the SM limit, $\delta g_{1,z}= \delta \kappa_\gamma = \lambda_z = 0$. 
Non-zero aTGCs are effectively generated in models with new heavy particles, after the latter are integrated out at low energies. 
Starting from a minimally coupled renormalizable UV theory, only $\delta g_{1,z}$ is generated at tree level \cite{Pomarol:2013zra,Biekoetter:2014jwa},  however at a loop level all 3 aTGCs can be generated with arbitrary coefficients depending on the matter content of the theory.   
Given many possible forms that physics  beyond the SM (BSM) could take, we think it is important to pursue a bottom-up approach in which as few assumptions as possible about the BSM sector are made. 
Consequently, in this paper we will always allow all 3 aTGCs to be present simultaneously with arbitrary coefficients. 
We also note that with this model-independent approach our results can be readily translated to any different basis of $D=6$ operator, which is in general not true if arbitrary assumptions about the aTGCs are made.

Non-zero aTGCs affect experimental observables, such as the total and differential WW pair production cross section in high-energy colliders. 
Precision measurements of these {quantities} at the LEP-2  $e^+e^-$ collider allow one to constrain these coefficients \cite{Schael:2013ita} (see e.g. \cite{Bian:2015zha} for future collider prospects).
Ref.~\cite{Falkowski:2014tna} performed a simultaneous fit of the three aTGCs to the LEP-2 data at $\cO(\Lambda^{-2})$ in the EFT.
That analysis revealed that robust limit on $\delta g_{1,z}$ and $\lambda_z$ are very weak, due to an accidental approximate flat direction of the fit for
$\delta g_{1,z} \approx -\lambda_z$.
Along this flat direction,  $\delta g_{1,z}$ and $\lambda_z$ of order $\sim 1$ are allowed by the LEP-2 data while the constraints on the orthogonal direction are at the $\cO(0.1)$ level. 

In principle, the flat direction can be lifted by precision measurements of the WW and WZ differential production cross sections at the LHC. 
Unfortunately, a robust EFT analysis of these data has not yet been presented by the experimental collaborations, 
and is difficult to perform with  theorist-level tools using the publicly available information.  
Meanwhile, it has been pointed out that an independent set of observables - the LHC Higgs data - can also lead to strong constraints on the aTGCs \cite{Corbett:2013pja,Pomarol:2013zra,Masso:2014xra,Ellis:2014jta,Dumont:2013wma}.
However, these analyses are not completely general from the EFT point of view:
the quadratic contributions in the aTGCs to the Higgs observables, formally of $\cO(\Lambda^{-4})$, are included, and/or not all possible $D$=6 operators affecting the Higgs observables are taken into account.  
We amend it in this letter.  

We derive constraints on the aTGCs  from the  combined LHC Higgs data and LEP-2 WW data sets. 
In our analysis, all $D$=6 operators affecting Higgs couplings to matter and gauge boson self-couplings are allowed to be simultaneously present  with arbitrary coefficients, assuming minimal flavor violation (MFV) \cite{DAmbrosio:2002ex}.
In the Higgs basis~\cite{HXSWGbasis} these parameters are \cite{Falkowski:2015fla}: 
\beq 
\label{eq:ind} 
 \delta c_z, \ c_{zz}, \   c_{z \Box}, \ \ c_{\gamma \gamma}, \ c_{z \gamma},  \  c_{gg},  \  \delta y_u, \   \delta y_d,  \ \delta y_e,  \ \lambda_z .
\eeq
Note that the dependence of the EFT cutoff $\Lambda$ is included in the operator coefficients.
The relation of these parameters to the interaction terms in the effective Lagrangian, as well as the relation to the aTGCs, can be found in Ref.~\cite{HXSWGbasis}.
Furthermore, we only take into account linear corrections in the Wilson coefficients, thus working consistently at the $\cO(\Lambda^{-2})$ in the EFT expansion. 
Note that, since different bases of $D=6$ operators in the literature  differ by  $\cO(\Lambda^{-4})$ terms corresponding to $D>6$ operators, only results obtained consistently at $\cO(\Lambda^{-2})$ are basis-independent~\footnote{We thank Michele Redi for this comment.}.
For the WW data,  we use the measured total and differential $e^+ e^- \to W^+ W^-$ cross sections different center-of-mass energies listed in Ref.~\cite{Schael:2013ita}.
{These cross sections depend on a number of EFT parameters in addition to the aTGCs, in particular on the ones inducing corrections to Z and W propagators and couplings to electrons. 
However,  given the  model-independent electroweak precision constraints~\cite{Efrati:2015eaa}, these measurements can effectively constrain  3 linear combinations of  Wilson coefficients of $D$=6 operators that correspond to the aTGCs~\cite{Falkowski:2014tna}. }
We use this dependence to construct the 3D  likelihood function 
$\chi^2_{WW} (\delta g_{1,z}$, $\delta \kappa_\gamma$, $\lambda_z)$.
For the LHC Higgs data,  we use the signal strength observables, that is, the ratio between the measured Higgs yield and its SM prediction $\mu \equiv (\sigma \times \textrm{BR})/(\sigma \times \textrm{BR})_{\textrm{SM}}$, listed in \tref{exp}, separated according to the final state and the production mode.
%
The effect of  $D$=6 operators on $\mu$ was calculated for each channel and production mode in Ref.~\cite{Falkowski:2015fla} and independently cross-checked here. 
After imposing electroweak precision constraints,  9 linear combinations of $D$=6 operators can affect $\mu$ in an observable way \cite{Corbett:2012ja,Pomarol:2013zra}. 
The crucial point is that 2 of these combinations correspond to the aTGCs $\delta g_{1,z}$, $\delta \kappa_\gamma$.
Therefore, the likelihood function constructed from LHC Higgs data, $\chi^2_{h}(\delta g_{1,z}, \delta \kappa_\gamma, \dots)$, may lead to additional constraints on aTGCs. 
Indeed, combining the likelihoods $\chi^2_{\rm comb.} = \chi^2_{h} + \chi^2_{WW}$ we obtain strong constraints on the aTGCs at the level of $\cO(0.1)$. Namely,  we obtain the likelihood for the three variables only: $\delta g_{1,z}$, $\delta \kappa_\gamma$ and $\lambda_z$, 
after minimizing at each point the combined likelihood with respect to the remaining seven Wilson coefficients. We find the following central values, 1~$\sigma$ errors, and the correlation matrix for the aTGCs:
\beq\begin{split}
\label{eq:constraints}
\bvec 
\delta g_{1,z} \\ \delta \kappa_\gamma \\ \lambda_z 
\evec   &= \bvec 0.043  \pm  0.031 \\ 0.142 \pm 0.085 \\  -0.162 \pm 0.073 \evec, \\ 
\rho &= \left ( \ba{ccc} 
1 & 0.74 & -0.85 \\ 
0.74 & 1 & -0.88 \\
-0.85 & -0.88 & 1 
\ea \right ).
\end{split}\eeq 
These constraints hold in any new physics scenario predicting approximately flavor blind coefficients of $D$=6 operators and in which $D> 6$ operators are subleading. 
\aref{fit} contains a technical description of our fit and the constraints for all the 10 combinations of Wilson coefficients entering the analysis. 
They are given in different bases for reader's convenience.

Let us discuss here qualitatively the most important elements of our fit.
Higgs data are sensitive to $\delta g_{1,z}$ and $\delta \kappa_\gamma$ primarily via their contribution to electroweak Higgs production channels.
However, only 1 combination of these 2 aTGCs is strongly constrained, while the bound on the  direction $\delta \kappa_\gamma \approx 3.8 \delta g_{1,z}$ is very weak.
Analogously, as already discussed, also LEP-2 bounds present an approximate blind direction.
This is illustrated in \fref{TGCbound}, where the WW and Higgs constraints in the $\delta g_{1,z}$--$\delta \kappa_\gamma$ plane are shown separately~\footnote{The bounds from WW data alone slightly depend on the scale at which the SM couplings $g$ and $g'$ are evaluated. Working at LO, these are higher-order effects and thus any scale is equally valid. Differently than in Ref.~\cite{Falkowski:2014tna}, here we extract those couplings from $G_F, m_Z$ and $\alpha_{em}(m_Z)$ and then run them up to the LEP-2 energy $\sqrt{s} \approx 200$~GeV. 
The resulting values are $g \approx 0.645$ and $g' \approx 0.357$. The dependence of the combined Higgs and WW fit on this choice is instead very small.}. 
Since the flat directions are nearly orthogonal,  combining LHC Higgs and LEP-2 WW data leads to the non-trivial constraints on aTGCs displayed in \eref{constraints}.

\begin{figure}[t]
\includegraphics[width=0.45 \textwidth]{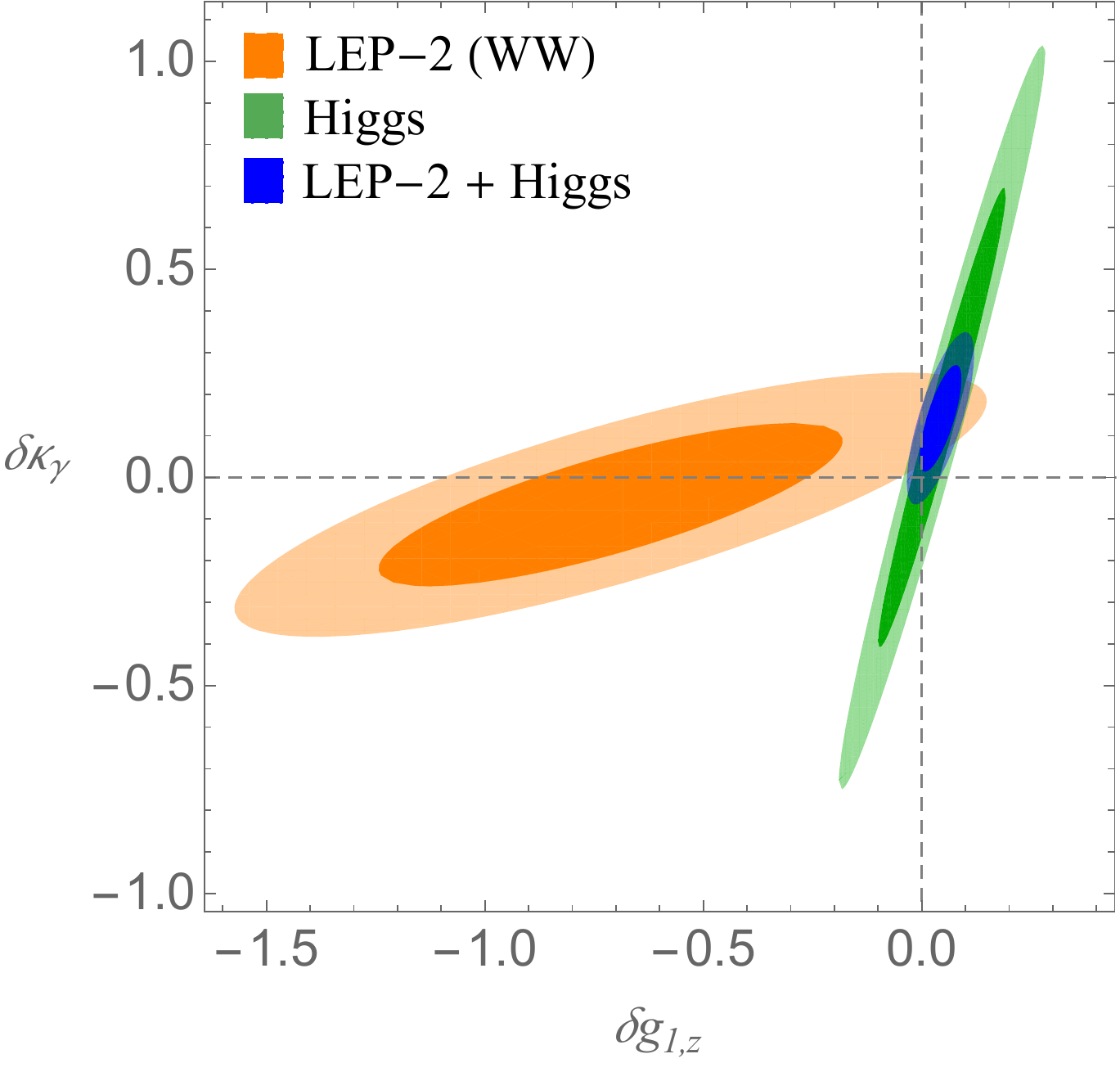}
\quad 
\caption{Allowed 68\% and 95\% CL region in the $\delta g_{1,z}$-$\delta \kappa_\gamma$ plane after considering LEP-2 $WW$ production data (TGC), Higgs data, and the combination of both datasets.
}
\label{fig:TGCbound}
\end{figure} 

One could further strengthen the constraints on aTGCs by considering the process of single on-shell W boson production in association with an electron and a neutrino ($e^+e^-\to WW^*\to W e \nu $)~\cite{Schael:2013ita}, as in Ref.~\cite{Falkowski:2014tna}. 
That process probes mostly $\delta \kappa_\gamma$ but it also affects limits on the remaining aTGCs due to the highly correlated nature of the constraints from  WW and Higgs data.
Indeed, we find that adding single W data to the combined likelihood roughly halves the confidence  intervals for the aTGCs: 
$\delta g_{1,z} = 0.017 \pm 0.023$, $\delta \kappa_\gamma = 0.047 \pm 0.034$, $\lambda_z  = -0.089 \pm 0.042$. 
However, we choose to highlight the more conservative result in \eref{constraints} as we consider it more robust.
The reason is that the experimental extraction of the single W cross section from fiducial measurements could be altered in a non-trivial way in the presence of the aTGC $\delta \kappa_\gamma$, which affects the photon t-channel contribution to the production amplitude.  
A more careful analysis is needed to render the single W constraint  more robust.    

In the following we discuss whether the assumptions employed in our analysis can be relaxed without conflicting experimental data and, if yes, how this affects our results.

We begin by considering the possible impact of  $D$=8 operators, contributing at $\cO(\Lambda^{-4})$.
In their presence one obtains a more complicated structure of aTGCs going beyond the 3-parameter characterization in \eref{atgc}.   
This is  likely to open new flat directions in the fit,  if the coefficients of the new aTGCs are allowed to be arbitrarily large. If the EFT expansion is valid, then the new contributions are  suppressed by $v^2 / \Lambda^2$ and therefore they are subleading with respect to the $3$ aTGCs taken into account in our fit.
However, since the experimental precision at the LHC is currently moderate, $\cO(20\%)$ at best, only higher-dimensional operators with  $\Lambda \lesssim $~few hundred~GeV can be constrained by Higgs physics. 
For such a low $\Lambda$ it is  not a priori obvious that the $D$=8 operators are subleading.
One way to estimate their effect is to include in the analysis corrections to Higgs and WW observables that are {\em quadratic} in the Wilson coefficients of $D$=6 operators, as they are also of $\cO(\Lambda^{-4})$.
If the constraints on the aTGCs are severely affected by including the quadratic contributions, that would signal a potential sensitivity to $D$=8 operators~\footnote{Keeping the quadratic terms while neglecting  $D$=8 operators can be justified for certain classes of UV completions of the EFT \cite{Biekoetter:2014jwa}.}.  
In fact, constraints from Higgs or from  WW data alone are completely changed after including the quadratic terms. 
However, the {\em combined} data are only moderately sensitive.
Once the quadratic contributions are included we find the constraints 
$\delta g_{1,z} = 0.032^{+0.043}_{-0.035}$,  $\delta \kappa_\gamma = 0.073^{+0.085}_{-0.075}$, $\lambda_z = -0.098^{+0.058}_{-0.065}$.
The confidence intervals are shifted at the level of  1~$\sigma$, however qualitative conclusions concerning the strength of the constraints on the aTGCs remain unchanged. 
 
The sensitivity to $D$=8 operators is particularly relevant for the Higgs production in association with a W or Z boson ($\sigma_{Vh}$), especially {for} large invariant mass of the $Vh$ system $m_{Vh}$ \cite{Biekoetter:2014jwa}.  
Although our  results change very little with the inclusion of the quadratic contributions, we note that they can be significant around the best-fit point. This should be examined with care since the validity of our EFT expressions for $\sigma_{Vh}$ is essential for lifting the flat direction of the LEP-2 data and obtaining strong constraints on aTGCs. We note that {the} EFT interpretation of Higgs searches could be made more robust if the $m_{Vh}$ distributions were available. 
Indeed, we find that $\sigma_{Vh}~(m_{Vh} < 400$ GeV) has a similar (reduced) sensitivity to linear (quadratic) terms. Therefore, assuming experimental measurements of $\mu_{Vh}$ do not significantly alter with this cut, we find that the TGC bounds remain unchanged, with a reduced sensitivity to higher-dimensional operators.
 
Furthermore, for low values of the EFT scale $\Lambda$, the presence of  $D>6$ operators whose contribution is larger or comparable to that of  $D$=6 operators could affect the Z-pole constraints on the latter (see also \cite{Berthier:2015oma}).
This in turn may affect the  per mille level constraints on the $Z$ and $W$ couplings to fermions assumed in our analysis. 
However, since the constraints we obtained on the aTGC are at the $\mathcal{O}(10\%)$ level, we do not expect this effect to qualitatively change our results unless  the constraints on the  $Z$ and $W$ coupling are relaxed to a similar level.

The question of the effect of the CP odd operators is closely related to the discussion  above. 
Our constraints are  based on Higgs signal strength observables, and on the total cross section and angular distribution of WW production. 
One can show that these are affected by CP odd operators only at  $\cO(\Lambda^{-4})$, as the CP violating contributions to the amplitude do not interfere with the SM contributions.
Therefore, their effect on the aTGCs bounds is of the same order as that of $D$=8 operators. 
 
The NLO EFT corrections to Higgs observables can be divided into 2 groups. 
The QCD corrections, $\cO\left({\alpha_s v^2 \over 4 \pi \Lambda^2} \right )$, do not affect $e^+ e^- \to W^+ W^-$ but they can contribute corrections as large as  $\cO(100\%)$ to Higgs processes.  
Fortunately, in the Higgs signal strength observables, which  involve a ratio of the observed to the predicted SM event rate, these large corrections are similar in the SM and new physics case and therefore they largely cancel (see e.g. \cite{Gori:2013mia} for the discussion in the context of the gluon fusion production process). 
The electroweak corrections, $\cO\left({\alpha_{\rm ew} v^2 \over 4 \pi \Lambda^2} \right )$, are in general non-factorizable. 
Some of these corrections correspond to a redefinition of the EFT parameters describing the Higgs couplings. 
For example, the  effect of the recently calculated NLO EFT corrections to the $h \to \gamma \gamma$ decay \cite{Hartmann:2015oia,Ghezzi:2015vva,Hartmann:2015aia} is to  replace the EFT parameter $c_{\gamma \gamma}$ with a linear combination of the renormalized $c_{\gamma \gamma}$ and other EFT parameters. 
This has no impact whatsoever on our determination of the aTGCs. 
More generally, {\em logarithmically enhanced} NLO corrections respect the structure of the tree-level Lagrangian and do not affect the relations between different couplings in the effective Lagrangian that are assumed in the fit. 
On the other hand, {\em finite} NLO corrections may affect these relations. 
However, they  are expected to be small, relative $\cO(10\%)$, and given the current experimental precision of LHC Higgs observables they should not affect the analysis in any significant way. 
{Finally, the NLO corrections to electroweak precision observables may affect bounds on certain operators, however the resulting feedback on WW and Higgs observables should again be negligible. }

Let us finally discuss the case in which no flavor symmetry is assumed.
The main effects impacting the fit are: 1) possible new operators affecting EW Higgs production and  decay,   2) possible large values of the Yukawa couplings to light fermions.

As  for the 1st point, in the EFT approach the coefficients of $hZff$ and $hWff'$ interactions are directly related to vertex corrections to the corresponding  $Zff$ and $Wff'$ interactions. 
To estimate their possible effect, we use the results of  a recent analysis of electroweak precision data in the flavor general EFT \cite{Efrati:2015eaa}. 
While most such terms are constrained with percent, or better, precision, vertex corrections to the couplings involving light quarks can be $\mathcal{O}(10\%)$.
This would weaken (though not dramatically) the TGC bounds obtained in the flavor blind case. However, the situation improves after taking into account also data from single $Z$ and $W$ Drell-Yan production at the LHC, where deviations from the SM are constrained at a few-percent level \cite{Chatrchyan:2014mua}. This is because the combination of $Zqq$ and $Wqq'$ couplings entering these processes is very similar to the one affecting the Higgs production cross section. 
 Including the constraints from Ref.~\cite{Chatrchyan:2014mua}, we find that the TGC limits in Eq.~\eqref{eq:constraints} hold with negligible modifications: the biggest effect is on $\delta g_{1,z}$ where the constraints are 15\% weaker.

Concerning  the 2nd point, the limits on the Higgs couplings in any current analysis crucially depend on assuming the MFV scenario, where the modifications of all Higgs Yukawa couplings can be related  to just 3 parameters $\delta y_{t,b,\tau}$ describing the Higgs couplings to the 3rd generation fermions. 
Going beyond the MFV scenario, all Yukawa couplings become free parameters. 
Allowing the Higgs coupling to muons to be a free parameter has no appreciable consequences for aTGCs because the $h\to\mu\mu$ data, which are included in our fit, constrain $\delta y_\mu$. 
On the other hand, data on tagged Higgs decays to quarks ($c,s,d,u$) are currently not  available. 
Therefore the only sensitivity of our set of observables to these couplings is via the modification of the total Higgs width and their one-loop contribution to the $ggH$ coupling {(see however \cite{Goertz:2014qia}).}
Thus, our TGC bounds  remain {unchanged} if we allow for flavor independent couplings, as long as they are not much bigger than the SM value. 
The situation when they are allowed to be much bigger than their SM values is more complicated (see e.g. Refs.~\cite{Perez:2015lra,Perez:2015aoa,Brivio:2015fxa}). 
Notice that large values are not incompatible with the EFT framework (as long as no flavor symmetry is assumed), which in principle predicts natural values of order $v^2/\Lambda^2$ that easily exceed the small Yukawas. 
Remarkably, even when light Yukawa couplings are as large as the bottom Yukawa (which would almost double the total Higgs width), our TGC bounds given in Eq.~\eqref{eq:constraints} qualitatively hold.

As discussed earlier, in this analysis we took into account only LEP-2 data on $e^+ e^- \to W^+ W^-$, and we ignored the Tevatron and LHC data on $WW$ and $WZ$ production in proton-(anti-)proton collisions. 
Including the latter could lead to better limits on the EFT parameters. 
However, to obtain robust constraints on aTGCs from hadron collider, the current analysis strategies need to be improved.
Unlike $e^+ e^-$ collisions at LEP, hadron collisions probe a wide range of energies, part of which may be beyond the validity regime of the EFT approach.  
The TGC analyses should therefore restrict the range of center-of-mass energies of partonic collisions from which the constraints are derived to be below the EFT cut-off $\Lambda$. 
Since the  cut-off is of course not known a-priori, the results should be quoted in function of $\Lambda$ (as also proposed in Ref.~\cite{Biekoetter:2014jwa} in  the context of $VH$ associated production).   
Next, the analyses should allow all 3 aTGCs to be present simultaneously, and correlation matrix for the  constraints on different parameters should be given. Finally, the analysis should be performed consistently at $\cO(\Lambda^{-2})$ in the EFT expansion, and the effects of neglecting or not $\cO(\Lambda^{-4})$ contributions should be quantified.   

As a final comment, we note that the tight bounds we obtain via the combination of LEP-2 WW and LHC Higgs data strongly constrain deviations in the $h \to 4 \ell$ distributions,
 which will be investigated in the LHC Run-2.
These decays can be described experimentally through a set of pseudo-observables \cite{Gonzalez-Alonso:2014eva}, which can then be matched to the $D$=6 operators in the EFT at tree-level \cite{Gonzalez-Alonso:2015bha}.
The strong bounds we obtain on the pseudo-observables from our fit are very similar to those presented in Ref.~\cite{Gonzalez-Alonso:2015bha} using only LEP2 data with $\lambda_z = 0$. 
Therefore, to a good approximation, the analysis performed in that work for such specific case holds now in full generality. In particular, the very strong bounds on the contact terms $\epsilon_{Z \ell_{L,R}}$ imply small deviations in the $h \to 4 \ell$ spectrum~\cite{Gonzalez-Alonso:2015bha}.

To conclude, by working at $\cO(\Lambda^{-2})$ in the EFT and under the MFV assumption, we obtained strong and model-independent bounds on the aTGCs via the combination of LEP-2 WW and LHC Higgs signal-strength data. The combination of the two datasets lifts the flat direction affecting each of them taken separately, thus showing the importance of performing global analysis in the EFT framework. Combined with the $W$- and $Z$-pole observables analysis of Ref.~\cite{Efrati:2015eaa}, the results of this work can be used to set strong constraints on a wide class of possible new physics scenarios.

\section*{Acknowledgements}
 
We thank Gino Isidori for useful discussions. AF is supported by the ERC Advanced Grant Higgs@LHC. M.G.-A. is grateful to the LABEX Lyon Institute of Origins (ANR-10-LABX-0066) of the Universit\'e de Lyon for its financial support within the program ANR-11-IDEX-0007 of the French government. A.G. and D.M. are supported in part by the Swiss National Science Foundation (SNF) under contract 200021-159720.

\appendix
\renewcommand{\theequation}{\Alph{section}.\arabic{equation}}  

\section{Fit results}
\label{app:fit}

In the SM extended by $D$=6 operators, assuming MFV,  there are 9 combinations of Wilson coefficients that  affect the Higgs signal strength measured at the LHC and are weakly constrained by electroweak precision tests.
Furthermore, to describe electroweak gauge bosons pair production, one more independent combination is needed.  
In the Higgs basis~\cite{HXSWGbasis} these 10 parameters are listed in eq.~\eqref{eq:ind}.
Their relation to the interaction terms in the effective Lagrangian can be found in Ref.~\cite{HXSWGbasis}. 
We constrain these parameters using the available LHC Higgs data  and WW data, as described above \eref{constraints}.
\begin{table}[t]
\begin{center}
\renewcommand*{\arraystretch}{1.2} 
\begin{tabular}{|c|l|l|c|c|}
\hline
Channel  & $\mu_{\rm{ATLAS}}$ & $\mu_{\rm{CMS}}$ & Production & Ref. 
\\ \hline 
$\gamma \gamma$ & $1.17^{+0.28}_{-0.26}$ &	$1.12^{+0.25}_{-0.22}$	& cats.   &  \cite{Aad:2014eha,Khachatryan:2014ira}
\\ \hline 
$Z \gamma$ & $2.7^{+4.6}_{-4.5}$ &	$-0.2^{+4.9}_{-4.9}$	& total   &  \cite{atlascoup,Chatrchyan:2013vaa}
\\ \hline 
$Z Z^*$ & $1.46^{+0.40}_{-0.34}$ &	$1.00^{+0.29}_{-0.29}$	& 2D   &  \cite{Aad:2014eva,Khachatryan:2014jba}
\\ \hline 
$W W^*$ & $1.18^{+0.24}_{-0.21}$ &	$0.83^{+0.21}_{-0.21}$	& 2D   &  \cite{ATLAS:2014aga,Khachatryan:2014jba}
\\ \cline{2-5} 
& $2.1^{+1.9}_{-1.6}$ &	-	& Wh   &  \cite{Aad:2015ona}
\\ \cline{2-5} 
& $5.1^{+4.3}_{-3.1}$ &	-	& Zh   &  \cite{Aad:2015ona}
\\ \cline{2-5} 
  &	-	&	$0.80^{+1.09}_{-0.93}$ & Vh  & \cite{Khachatryan:2014jba}
\\ \hline 
$\tau \tau$ &	$1.44^{+0.42}_{-0.37}$	& $0.91^{+0.28}_{-0.28}$ & 2D   &  \cite{Aad:2015vsa,Khachatryan:2014jba}
\\ \cline{2-5} 
&-& $0.87^{+1.00}_{-0.88}$ & Vh   &  \cite{Khachatryan:2014jba}
\\ \hline 
$b b$  & $1.11^{+0.65}_{-0.61}$ &	-	& Wh   &  \cite{Aad:2014xzb}
\\ \cline{2-5} 
  & $0.05^{+0.52}_{-0.49}$ &	-	& Zh  &  \cite{Aad:2014xzb}
\\ \cline{2-5} 
  &	-	&	$0.89^{+0.47}_{-0.44}$	&	Vh	&	\cite{Khachatryan:2014jba}
\\ \cline{2-5} 
  &	-	&	$2.8^{+1.6}_{-1.4}$	&	VBF	&	\cite{Khachatryan:2015bnx}
\\ \cline{2-5} 
  & $1.5^{+1.1}_{-1.1}$ &	$1.2^{+1.6}_{-1.5}$	& tth   &  \cite{Aad:2015gra,Khachatryan:2015ila}
\\ \hline 
 $\mu \mu$  & $-0.7^{+3.7}_{-3.7}$ &	$0.8^{+3.5}_{-3.4}$	& total   &  \cite{atlascoup,Khachatryan:2014aep}
\\ \hline 
 multi-$\ell$  & $2.1^{+1.4}_{-1.2}$ &	$3.8^{+1.4}_{-1.4}$	& tth   &  \cite{atlasml,Khachatryan:2014qaa}
\\ \hline 
\end{tabular}
\quad 
\end{center}
\caption{
\label{tab:exp}
The LHC Higgs results used in the fit.
2D stands for the likelihood functions in the plane $\mu_{\rm ggh+tth}$-$\mu_{\rm VBF+Vh}$, whereas in the diphoton channel (cats.) we use the five-dimensional likelihood function in the space spanned by $(\mu_{ggh}, \mu_{tth}, \mu_{\rm VBF}, \mu_{Wh},\mu_{Zh})$. Notice that in these two cases $\mu$ is quoted for illustration only, since more information is included in the analysis. Correlations among different production classes in this table are ignored. See Ref.~\cite{Falkowski:2015fla} for a more detailed discussion of our Higgs dataset.}
\end{table}
In the Gaussian approximation near the best fit point we find the following constraints: 
\beq
\label{eq:fit}
\bvec \delta c_z \\  c_{zz} \\  c_{z\Box} \\  c_{\gamma \gamma} \\ c_{z\gamma} \\  c_{gg} \\  \delta y_u  \\ \delta y_d \\  \delta y_e \\ \lambda_z   \evec  = 
\bvec  -0.02 \pm 0.17 \\  0.69 \pm  0.42 \\ -0.32 \pm 0.19 \\  0.009 \pm 0.015  \\  0.002 \pm 0.098  \\ -0.0052 \pm 0.0027 \\ 0.57 \pm 0.30 \\ -0.24 \pm 0.35  \\ -0.12 \pm 0.20 \\ -0.162 \pm 0.073  \evec, 
\eeq 
where the uncertainties correspond to $1 \sigma$.  
The correlation matrix is given by 
\begin{equation}
\scriptsize
\label{eq:rho} 
\! \! 
\left ( \begin{array}{cccccccccc}
 1 & -.04  & -.21  & -.76 & -.15 & .15 & .12 & .88 & .71 & -.22  \\
\cdot&1 & -.96 & .37  & .19 &  .03 & .04 & -.12 & -.31 & -.88  \\
\cdot&\cdot&1 & -.17 & -.10 & -.07 & -.06 & -.10 & .12 & .93   \\
\cdot&\cdot&\cdot&1 & .20 & -.12  & -.07 & -.79 & -.74 & -.13   \\
\cdot &\cdot&\cdot&\cdot&1 & -.01 & -.01 & -.15 & -.18 & -.10  \\
\cdot&\cdot&\cdot&\cdot&\cdot&1 & -.87 & .26 & .17 & -.07   \\
\cdot&\cdot&\cdot&\cdot&\cdot&\cdot&1 & .13 & .11 & -.06  \\
\cdot&\cdot&\cdot&\cdot&\cdot&\cdot&\cdot&1 & .81 & -.11  \\
\cdot&\cdot&\cdot&\cdot&\cdot&\cdot&\cdot&\cdot&1 & .09 \\ 
\cdot&\cdot&\cdot&\cdot&\cdot&\cdot&\cdot&\cdot&\cdot&1 
\end{array} \right ).
\end{equation}
%
To translate these results into constraints on aTGCs in \eref{constraints}, one needs
the relation between  $\delta \kappa_\gamma$ and $\delta g_{1,z}$ and the Higgs basis parameters of \eref{ind} \cite{HXSWGbasis}:
\bea
	\delta g_{1,z} &=& \frac{1}{2(g^2 - g^{\prime 2})} \left[ - g^2 (g^2 + g^{\prime 2}) c_{z\Box} - g^{\prime 2} (g^2 + g^{\prime 2}) c_{zz} + \right. \nonumber \\
	&&\quad + \left. e^2 g^{\prime 2} c_{\gamma\gamma} +g^{\prime 2} (g^2 - g^{\prime 2} ) c_{z\gamma} \right]~, \nonumber\\
	\delta \kappa_\gamma &=& - \frac{g^2}{2} \left( c_{\gamma\gamma} \frac{e^2}{g^2 + g^{\prime 2}} + c_{z\gamma} \frac{g^2 - g^{\prime 2}}{g^2 + g^{\prime 2}} - c_{zz} \right)~.
	\label{eq:TGCdependent}
\eea

{In the rest of the appendix we} translate the results in  \eref{fit} to different  bases of $D$=6 operators used in the literature:  the so-called  \emph{Warsaw}, {\em SILH'}, and {\em HISZ} basis. 
In each case, we assume the Wilson coefficients respect MFV,  and we restrict to the 10-dimensional subspace of the Wilson coefficients that affects Higgs and WW observables, but in which  the  LEP-1 $Z$-pole observables{, constrained at the permil level,} are not affected.  
The relation map between the parameters in \eref{ind} and the Wilson coefficients in these bases  can be found in Ref.~\cite{HXSWGbasis}, 
while the directions in the parameter space affecting electroweak precision observables are characterized in Ref.~\cite{Efrati:2015eaa}. 

\subsection*{Translation to Warsaw basis}

We consider the Warsaw basis of Ref.~\cite{Grzadkowski:2010es}, up to small modifications defined in Ref.~\cite{HXSWGbasis}.  
As before, we assume the Wilson coefficients of $D$=6 operators respect MFV.   
We  use the normalization and notation of Ref.~\cite{HXSWGbasis}, except that  we  rescale the Wilson coefficients of the Yukawa operators as $c_f \to  {\sqrt{2} m_f \over v} \hat c_f$.
LEP-1 electroweak precision observables are not affected if 
\bea
\label{eq:EWPT_chats}
& 
c_{H\ell}'   = c_{Hq}'   = -g^2 c_{WB}  +  {g^2 \over  g'{}^2} c_T, 
\nnl & 
c_{H\ell}  =  {1 \over 2} c_{He} =   - 3 c_{Hq} = - {3 \over 4} c_{Hu} = {3 \over 2} c_{Hd} = c_T.  
\eea   
We impose these constraints in our fit.
Out of the remaining Wilson coefficients, only  10 affect Higgs and WW observables at the tree level.  
They are  constrained as:
\beq
\left(
\begin{array}{rl}
 c_{H} & = 0.11 \pm 0.15 \\
  c_{T} & = 0.034 \pm 0.021 \\
 c_{WB} & = 0.34 \pm 0.20 \\
 c_{WW} & = 0.69 \pm 0.43 \\
 c_{BB} & = 0.69 \pm 0.42 \\
 c_{GG} & = -0.0052 \pm 0.0027 \\
 \hat c_{u} & = 0.65 \pm 0.32 \\
 \hat  c_{d} & = -0.16  \pm 0.23 \\
 \hat  c_{e} & = -0.03 \pm 0.13 \\
  c_{3W} & = 0.63 \pm 0.29  \\
\end{array}
\right)
~~~,
\eeq
with the correlation matrix:
\beq
\label{eq:CorrMatrixWB}
\scriptsize
\rho = \left(
\begin{array}{cccccccccc}
 1 & .51  & .38  & 0.43 & .34 & -.11 & .37 & -.62 & -.01 & .16  \\
\cdot&1 & .97 & .94  & .96 &  .00 & .22 & -.13 & -.17 & .79  \\
\cdot&\cdot&1 & .97 & .97 & .03 & .16 & .01 & -.16 & .88   \\
\cdot&\cdot&\cdot&1 & .89 & .03  & .18 & -.01 & -.16 & .87   \\
\cdot &\cdot&\cdot&\cdot&1 & .03 & .14 & .01 & -.15 & .84  \\
\cdot&\cdot&\cdot&\cdot&\cdot&1 & -.87 & .31 & .11 & .07   \\
\cdot&\cdot&\cdot&\cdot&\cdot&\cdot&1 & -.19 & .03 & .07  \\
\cdot&\cdot&\cdot&\cdot&\cdot&\cdot&\cdot&1 & .37 & .18  \\
\cdot&\cdot&\cdot&\cdot&\cdot&\cdot&\cdot&\cdot&1 & -.11 \\ 
\cdot&\cdot&\cdot&\cdot&\cdot&\cdot&\cdot&\cdot&\cdot&1 
\end{array} 
\right).
\eeq

\subsection*{Translation to SILH' basis}

We move to a variant of the SILH basis \cite{Giudice:2007fh} defined in Ref.~\cite{Pomarol:2013zra} and often referred to as SILH'.
Again,  for normalization of  operators and their Wilson coefficients we use Ref.~\cite{HXSWGbasis}, we assume MFV,  and  we  rescale the Wilson coefficients of the Yukawa operators as $s_f \to  {\sqrt{2} m_f \over v} \hat s_f$. 
In this basis, the relations due to LEP-1 electroweak precision observables are simpler: 
$s_T =  s_{\ell \ell} = s_{Hf} = s_{Hf}' = 0$, and $s_W + s_B =0$. 
This implies that, after including also LEP-1 data in a global analysis, the correlation matrix becomes block-diagonal to a very good accuracy.
The remaining 10 parameters affecting Higgs and WW observables are constrained as: 
\beq
\left(
\begin{array}{rl}
 s_{H} &= 0.02\pm 0.17 \\ 
 {1 \over 2}\left (s_{W} - s_B \right ) &= 0.37 \pm 0.30 \\
 s_{HW} &= -0.69 \pm 0.43 \\
 s_{HB} &= -0.68 \pm 0.42 \\
 s_{BB} &= 0.094 \pm 0.015 \\
 s_{GG} &= -0.0052 \pm 0.0027 \\
 \hat s_{u} &= 0.59 \pm 0.33 \\
\hat  s_{d} &= -0.23 \pm 0.22 \\
 \hat s_{e} &= -0.10 \pm 0.15 \\
  s_{3W} &= 0.63 \pm 0.29  
\end{array}
\right)~,
\eeq
with the correlation matrix: 
\beq
\label{eq:CorrMatrixSILHp}
\scriptsize
\rho = \left(
\begin{array}{cccccccccc}
 1 & .49  & -.05  & 0.03 & .76 & -.15 & .41 & -.63 & .19 & -.22  \\
\cdot&1 & -.85 & .73  & .65 &  -.06 & .25 & -.42 & -.27 & .59  \\
\cdot&\cdot&1 & .89 & -.38 & -.03 & -.07 & .18 & .39 & -.87   \\
\cdot&\cdot&\cdot&1 & -.28 & -.04  & -.03 & .11 & .36 & -.85   \\
\cdot &\cdot&\cdot&\cdot&1 & -.12 & .33 & -.68 & -.14 & .13  \\
\cdot&\cdot&\cdot&\cdot&\cdot&1 & -.88 & .30 & .06 & .07   \\
\cdot&\cdot&\cdot&\cdot&\cdot&\cdot&1 & -.22 & .10 & -.06  \\
\cdot&\cdot&\cdot&\cdot&\cdot&\cdot&\cdot&1 & .30 & .01  \\
\cdot&\cdot&\cdot&\cdot&\cdot&\cdot&\cdot&\cdot&1 & -.38 \\ 
\cdot&\cdot&\cdot&\cdot&\cdot&\cdot&\cdot&\cdot&\cdot&1 
\end{array} 
\right).
\eeq

\subsection*{Translation to HISZ basis}

Finally, we translate our results  into the language of the HISZ operator set \cite{Hagiwara:1993ck}, following  the conventions of Ref.~\cite{Corbett:2015ksa}  with $\Lambda = v$. 
We obtain 
\beq
\left(
\begin{array}{rl}
 f_{H,2} &= 0.03 \pm 0.34 \\ 
 f_{W} &= 0.64 \pm 0.46 \\
 f_{B} &= 2.11 \pm 1.33 \\
 f_{WW} &= -0.37 \pm 0.30 \\
 f_{BB} &= 0.36  \pm 0.29 \\
 f_{GG} &= 0.41 \pm 0.21 \\
 f_{u} &= -0.83 \pm 0.46 \\
f_{d} &=  0.32 \pm 0.31 \\
 f_{e} &=  0.14 \pm 0.20 \\
 f_{3W} &= -2.53 \pm 1.14  
\end{array}
\right)~,
\eeq
with the correlation matrix: 
\beq
\label{eq:CorrMatrixHISZ}
\scriptsize
\rho = \left(
\begin{array}{cccccccccc}
 1 & -.53 & .20  & -.49 & .47 & .15 & -.41 & .63 & -.19 & .22  \\
\cdot&1 & .56 & -.29  & .31 &  -.13 & .20 & -.22 & -.38 & -.85  \\
\cdot&\cdot&1 & -.91 & .91 & .00 & -.13 & .26 & .35 & -.79   \\
\cdot&\cdot&\cdot&1 & -.999 & -.06  & .25 & -.42 & -.27 & .59   \\
\cdot &\cdot&\cdot&\cdot&1 & .06 & -.24 & .40 & .27 & -.06  \\
\cdot&\cdot&\cdot&\cdot&\cdot&1 & -.88 & .30 & .06 & .07   \\
\cdot&\cdot&\cdot&\cdot&\cdot&\cdot&1 & -.22 & .10 & -.06  \\
\cdot&\cdot&\cdot&\cdot&\cdot&\cdot&\cdot&1 & .30 & .01  \\
\cdot&\cdot&\cdot&\cdot&\cdot&\cdot&\cdot&\cdot&1 & -.38 \\ 
\cdot&\cdot&\cdot&\cdot&\cdot&\cdot&\cdot&\cdot&\cdot&1 
\end{array} 
\right).
\eeq

{
\subsection*{$h \to 4\ell$ pseudo-observables}
}

{Here we report the bounds on the Higgs pseudo-observables relevant to $h\to4\ell$ decays, obtained via} a tree-level matching with the $D$=6 operators 
in the Higgs basis \cite{Gonzalez-Alonso:2015bha}. {At this level,} only five pseudo-observables are independent and the constraints we find are:
\beq
\scriptsize
\begin{split}
\label{eq:HPObounds}
\bvec \kappa_{ZZ}=0.85  \pm  0.17 \\ \epsilon_{Z \ell_L}=-0.0001 \pm 0.0078 \\  \epsilon_{Z \ell_R}=-0.025 \pm 0.015 \\  \kappa_{Z\gamma}=0.96 \pm 1.6 \\ \kappa_{\gamma\gamma}=0.88 \pm 0.19 \evec, \;
\rho = \left ( \ba{ccccc} 
1 & .72 & .60 & .19 & .83 \\
 \cdot & 1 & .35 & -.16 & .62 \\
 \cdot & \cdot & 1 & .02 & .47 \\
 \cdot & \cdot & \cdot & 1 & .20 \\
 \cdot & \cdot & \cdot & \cdot & 1 \\
\ea \right ).
\end{split}
\eeq

\section{Single $Z$ and $W$ Drell-Yan production}
\label{app:DY}

Using Madgraph5\_aMC@NLO~\cite{Alwall:2014hca} we compute the leading order (LO) contribution of {the} $D$=6 operators in the Higgs basis to the $Z$- and $W$-boson Drell-Yan production cross-section at 8~TeV in the flavor-general EFT finding:
\beq
\begin{split}
	\frac{\sigma_{\text{LO}} (pp \rightarrow Z)}{\sigma_{\text{SM,\,LO}} (pp \rightarrow Z)} =& \; 1 + 2.20\, \delta g^{Z u}_L - 1.01\, \delta g^{Z u}_R \\
	&- 1.89\, \delta g^{Z d}_L  + 0.34\, \delta g^{Z d}_R~, \\
\frac{\sigma_{\text{LO}}  (pp \rightarrow W)}{\sigma_{\text{SM,\,LO}} (pp \rightarrow W)} =&~ 1 + 1.73\, (\delta g^{Z u}_L - \delta g^{Z d}_L)~,
\label{eq:DYbsm}
\end{split}
\eeq
where $\sigma_{\text{SM,\,LO}} (pp \rightarrow Z) \approx 23.9 \text{ nb}$ and $\sigma_{\text{SM,\,LO}} (pp \rightarrow W) \approx 84.5 \text{ nb}$. From Ref.~\cite{Chatrchyan:2014mua}, 
we get the experimental constraints from 8~TeV data:
\beq
	\frac{\sigma_{\text{exp}} (pp \rightarrow Z \rightarrow \ell^+ \ell^-)}{\sigma_{\text{SM, NNLO}} (pp \rightarrow Z \rightarrow \ell^+ \ell^-)} = 1.02 \pm 0.05.
\eeq
\beq
	\frac{\sigma_{\text{exp}} (pp \rightarrow W \rightarrow \ell \nu)}{\sigma_{\text{SM, NNLO}} (pp \rightarrow W \rightarrow \ell \nu)} = 1.00 \pm 0.04.
\eeq
As explained in the main text, we assume that the NLO QCD corrections largely cancel in the BSM vs SM ratio of Eq.~\eqref{eq:DYbsm}, and that NLO EW corrections can be neglected. Taking into account that NP effects in leptonic Z decays are negligible at this level of precision~\cite{Efrati:2015eaa}, we use these experimental results to improve the bounds on the {$\delta g^{Vq}_{L,R}$ coefficients} obtained from LEP1 data in Ref.~\cite{Efrati:2015eaa}. 
 
These limits are used to constrain the extra contribution to the production modes VBF, $Wh$ and $Zh$ due to such anomalous $W$ and $Z$ couplings, which are given by
\bea
\frac{\delta \sigma_{VBF}}{\sigma_{VBF}^{SM}} &=& - 6.7 \delta g^{Zu}_L + 0.9 \delta g^{Zu}_R + 6.1 \delta g^{Zd}_L - 0.28 \delta g^{Zd}_R~,\nonumber\\
\frac{\delta \sigma_{ W h}}{\sigma_{ W h}^{SM}} &=& 28 \delta g^{Zu}_L - 28 \delta g^{Zd}_L ~,\nonumber\\
\frac{\delta \sigma_{ Z h}}{\sigma_{ Z h}^{SM}} &=&  31 \delta g^{Zu}_L - 14  \delta g^{Zu}_R - 23 \delta g^{Zd}_L + 4.3 \delta g^{Zd}_R ~.
\eea

\bibliographystyle{JHEP}
\bibliography{tgchiggs}

\providecommand{\href}[2]{#2}\begingroup\raggedright\begin{thebibliography}{10}

\bibitem{DeRujula:1991se}
A.~De~Rujula, M.~B. Gavela, P.~Hernandez, and E.~Masso {\em Nucl. Phys.} {\bf
  B384} (1992) 3--58.

\bibitem{Hagiwara:1993ck}
K.~Hagiwara, S.~Ishihara, R.~Szalapski, and D.~Zeppenfeld {\em Phys.Rev.} {\bf
  D48} (1993) 2182--2203.

\bibitem{Pomarol:2013zra}
A.~Pomarol and F.~Riva {\em JHEP} {\bf 1401} (2014) 151,
  [\href{http://arxiv.org/abs/1308.2803}{{\tt arXiv:1308.2803}}].

\bibitem{Biekoetter:2014jwa}
A.~Biekoetter, A.~Knochel, M.~KrŠmer, D.~Liu, and F.~Riva {\em Phys. Rev.} {\bf
  D91} (2015) 055029, [\href{http://arxiv.org/abs/1406.7320}{{\tt
  arXiv:1406.7320}}].

\bibitem{Schael:2013ita}
{\bf ALEPH, DELPHI, L3, OPAL, LEP Electroweak} Collaboration, S.~Schael et~al.
  {\em Phys.Rept.} {\bf 532} (2013) 119--244,
  [\href{http://arxiv.org/abs/1302.3415}{{\tt arXiv:1302.3415}}].

\bibitem{Bian:2015zha}
L.~Bian, J.~Shu, and Y.~Zhang \href{http://arxiv.org/abs/1507.02238}{{\tt
  arXiv:1507.02238}}.

\bibitem{Falkowski:2014tna}
A.~Falkowski and F.~Riva {\em JHEP} {\bf 1502} (2015) 039,
  [\href{http://arxiv.org/abs/1411.0669}{{\tt arXiv:1411.0669}}].

\bibitem{Corbett:2013pja}
T.~Corbett, O.~Eboli, J.~Gonzalez-Fraile, and M.~Gonzalez-Garcia {\em
  Phys.Rev.Lett.} {\bf 111} (2013) 011801,
  [\href{http://arxiv.org/abs/1304.1151}{{\tt arXiv:1304.1151}}].

\bibitem{Masso:2014xra}
E.~Masso {\em JHEP} {\bf 1410} (2014) 128,
  [\href{http://arxiv.org/abs/1406.6376}{{\tt arXiv:1406.6376}}].

\bibitem{Ellis:2014jta}
J.~Ellis, V.~Sanz, and T.~You {\em JHEP} {\bf 1503} (2015) 157,
  [\href{http://arxiv.org/abs/1410.7703}{{\tt arXiv:1410.7703}}].

\bibitem{Dumont:2013wma}
B.~Dumont, S.~Fichet, and G.~von Gersdorff {\em JHEP} {\bf 1307} (2013) 065,
  [\href{http://arxiv.org/abs/1304.3369}{{\tt arXiv:1304.3369}}].

\bibitem{DAmbrosio:2002ex}
G.~D'Ambrosio, G.~Giudice, G.~Isidori, and A.~Strumia {\em Nucl.Phys.} {\bf
  B645} (2002) 155--187, [\href{http://arxiv.org/abs/hep-ph/0207036}{{\tt
  hep-ph/0207036}}].

\bibitem{HXSWGbasis}
{\bf LHC Higgs Cross Section Working Group 2} Collaboration, LHCHXSWG2
  \href{http://arxiv.org/abs/LHCHXSWG-INT-2015-001
  cds.cern.ch/record/2001958}{{\tt LHCHXSWG-INT-2015-001
  cds.cern.ch/record/2001958}}.

\bibitem{Falkowski:2015fla}
A.~Falkowski \href{http://arxiv.org/abs/1505.00046}{{\tt arXiv:1505.00046}}.

\bibitem{Note1}
We thank Michele Redi for this comment.

\bibitem{Efrati:2015eaa}
A.~Efrati, A.~Falkowski, and Y.~Soreq {\em JHEP} {\bf 07} (2015) 018,
  [\href{http://arxiv.org/abs/1503.07872}{{\tt arXiv:1503.07872}}].

\bibitem{Corbett:2012ja}
T.~Corbett, O.~Eboli, J.~Gonzalez-Fraile, and M.~Gonzalez-Garcia {\em
  Phys.Rev.} {\bf D87} (2013) 015022,
  [\href{http://arxiv.org/abs/1211.4580}{{\tt arXiv:1211.4580}}].

\bibitem{Note2}
The bounds from WW data alone slightly depend on the scale at which the SM
  couplings $g$ and $g'$ are evaluated. Working at LO, these are higher-order
  effects and thus any scale is equally valid. Differently than in Ref.~\cite
  {Falkowski:2014tna}, here we extract those couplings from $G_F, m_Z$ and
  $\alpha _{em}(m_Z)$ and then run them up to the LEP-2 energy $\protect \sqrt
  {s} \approx 200$~GeV. The resulting values are $g \approx 0.645$ and $g'
  \approx 0.357$. The dependence of the combined Higgs and WW fit on this
  choice is instead very small.

\bibitem{Note3}
Keeping the quadratic terms while neglecting $D$=8 operators can be justified
  for certain classes of UV completions of the EFT \cite {Biekoetter:2014jwa}.

\bibitem{Berthier:2015oma}
L.~Berthier and M.~Trott {\em JHEP} {\bf 05} (2015) 024,
  [\href{http://arxiv.org/abs/1502.02570}{{\tt arXiv:1502.02570}}].

\bibitem{Gori:2013mia}
S.~Gori and I.~Low {\em JHEP} {\bf 1309} (2013) 151,
  [\href{http://arxiv.org/abs/1307.0496}{{\tt arXiv:1307.0496}}].

\bibitem{Hartmann:2015oia}
C.~Hartmann and M.~Trott {\em JHEP} {\bf 07} (2015) 151,
  [\href{http://arxiv.org/abs/1505.02646}{{\tt arXiv:1505.02646}}].

\bibitem{Ghezzi:2015vva}
M.~Ghezzi, R.~Gomez-Ambrosio, G.~Passarino, and S.~Uccirati
  \href{http://arxiv.org/abs/1505.03706}{{\tt arXiv:1505.03706}}.

\bibitem{Hartmann:2015aia}
C.~Hartmann and M.~Trott \href{http://arxiv.org/abs/1507.03568}{{\tt
  arXiv:1507.03568}}.

\bibitem{Chatrchyan:2014mua}
{\bf CMS} Collaboration, S.~Chatrchyan et~al. {\em Phys.Rev.Lett.} {\bf 112}
  (2014) 191802, [\href{http://arxiv.org/abs/1402.0923}{{\tt
  arXiv:1402.0923}}].

\bibitem{Goertz:2014qia}
F.~Goertz {\em Phys. Rev. Lett.} {\bf 113} (2014), no.~26 261803,
  [\href{http://arxiv.org/abs/1406.0102}{{\tt arXiv:1406.0102}}].

\bibitem{Perez:2015lra}
G.~Perez, Y.~Soreq, E.~Stamou, and K.~Tobioka
  \href{http://arxiv.org/abs/1505.06689}{{\tt arXiv:1505.06689}}.

\bibitem{Perez:2015aoa}
G.~Perez, Y.~Soreq, E.~Stamou, and K.~Tobioka
  \href{http://arxiv.org/abs/1503.00290}{{\tt arXiv:1503.00290}}.

\bibitem{Brivio:2015fxa}
I.~Brivio, F.~Goertz, and G.~Isidori
  \href{http://arxiv.org/abs/1507.02916}{{\tt arXiv:1507.02916}}.

\bibitem{Gonzalez-Alonso:2014eva}
M.~Gonzalez-Alonso, A.~Greljo, G.~Isidori, and D.~Marzocca {\em Eur.Phys.J.}
  {\bf C75} (2015), no.~3 128, [\href{http://arxiv.org/abs/1412.6038}{{\tt
  arXiv:1412.6038}}].

\bibitem{Gonzalez-Alonso:2015bha}
M.~Gonzalez-Alonso, A.~Greljo, G.~Isidori, and D.~Marzocca
  \href{http://arxiv.org/abs/1504.04018}{{\tt arXiv:1504.04018}}.

\bibitem{Aad:2014eha}
{\bf ATLAS} Collaboration, G.~Aad et~al. {\em Phys.Rev.} {\bf D90} (2014),
  no.~11 112015, [\href{http://arxiv.org/abs/1408.7084}{{\tt
  arXiv:1408.7084}}].

\bibitem{Khachatryan:2014ira}
{\bf CMS} Collaboration, V.~Khachatryan et~al. {\em Eur.Phys.J.} {\bf C74}
  (2014), no.~10 3076, [\href{http://arxiv.org/abs/1407.0558}{{\tt
  arXiv:1407.0558}}].

\bibitem{atlascoup}
{\bf ATLAS} Collaboration {\em ATLAS-CONF-2015-007} (2015).

\bibitem{Chatrchyan:2013vaa}
{\bf CMS} Collaboration, S.~Chatrchyan et~al. {\em Phys.Lett.} {\bf B726}
  (2013) 587--609, [\href{http://arxiv.org/abs/1307.5515}{{\tt
  arXiv:1307.5515}}].

\bibitem{Aad:2014eva}
{\bf ATLAS Collaboration} Collaboration, G.~Aad et~al. {\em Phys.Rev.} {\bf
  D91} (2015), no.~1 012006, [\href{http://arxiv.org/abs/1408.5191}{{\tt
  arXiv:1408.5191}}].

\bibitem{Khachatryan:2014jba}
{\bf CMS} Collaboration, V.~Khachatryan et~al.
  \href{http://arxiv.org/abs/1412.8662}{{\tt arXiv:1412.8662}}.

\bibitem{ATLAS:2014aga}
{\bf ATLAS Collaboration} Collaboration, G.~Aad et~al.
  \href{http://arxiv.org/abs/1412.2641}{{\tt arXiv:1412.2641}}.

\bibitem{Aad:2015ona}
{\bf ATLAS} Collaboration, G.~Aad et~al.
  \href{http://arxiv.org/abs/1506.06641}{{\tt arXiv:1506.06641}}.

\bibitem{Aad:2015vsa}
{\bf ATLAS} Collaboration, G.~Aad et~al.
  \href{http://arxiv.org/abs/1501.04943}{{\tt arXiv:1501.04943}}.

\bibitem{Aad:2014xzb}
{\bf ATLAS} Collaboration, G.~Aad et~al. {\em JHEP} {\bf 1501} (2015) 069,
  [\href{http://arxiv.org/abs/1409.6212}{{\tt arXiv:1409.6212}}].

\bibitem{Khachatryan:2015bnx}
{\bf CMS} Collaboration, V.~Khachatryan et~al.
  \href{http://arxiv.org/abs/1506.01010}{{\tt arXiv:1506.01010}}.

\bibitem{Aad:2015gra}
{\bf ATLAS} Collaboration, G.~Aad et~al.
  \href{http://arxiv.org/abs/1503.05066}{{\tt arXiv:1503.05066}}.

\bibitem{Khachatryan:2015ila}
{\bf CMS} Collaboration, V.~Khachatryan et~al.
  \href{http://arxiv.org/abs/1502.02485}{{\tt arXiv:1502.02485}}.

\bibitem{Khachatryan:2014aep}
{\bf CMS} Collaboration, V.~Khachatryan et~al.
  \href{http://arxiv.org/abs/1410.6679}{{\tt arXiv:1410.6679}}.

\bibitem{atlasml}
{\bf ATLAS} Collaboration {\em ATLAS-CONF-2015-006} (2015).

\bibitem{Khachatryan:2014qaa}
{\bf CMS} Collaboration, V.~Khachatryan et~al. {\em JHEP} {\bf 1409} (2014)
  087, [\href{http://arxiv.org/abs/1408.1682}{{\tt arXiv:1408.1682}}].

\bibitem{Grzadkowski:2010es}
B.~Grzadkowski, M.~Iskrzynski, M.~Misiak, and J.~Rosiek {\em JHEP} {\bf 1010}
  (2010) 085, [\href{http://arxiv.org/abs/1008.4884}{{\tt arXiv:1008.4884}}].

\bibitem{Giudice:2007fh}
G.~Giudice, C.~Grojean, A.~Pomarol, and R.~Rattazzi {\em JHEP} {\bf 0706}
  (2007) 045, [\href{http://arxiv.org/abs/hep-ph/0703164}{{\tt
  hep-ph/0703164}}].

\bibitem{Corbett:2015ksa}
T.~Corbett, O.~J.~P. Eboli, D.~Goncalves, J.~Gonzalez-Fraile, T.~Plehn, and
  M.~Rauch \href{http://arxiv.org/abs/1505.05516}{{\tt arXiv:1505.05516}}.

\bibitem{Alwall:2014hca}
J.~Alwall, R.~Frederix, S.~Frixione, V.~Hirschi, F.~Maltoni, et~al. {\em JHEP}
  {\bf 1407} (2014) 079, [\href{http://arxiv.org/abs/1405.0301}{{\tt
  arXiv:1405.0301}}].

\end{thebibliography}\endgroup

\end{document}